\documentstyle[12pt,epsfig]{article}

\begin{document}
\begin {center}
{\bf {\Large
Resonance nucleon scattering amplitude in the photonuclear reaction
} }
\end {center}
\begin {center}
Swapan Das  \\
{\it Nuclear Physics Division,
Bhabha Atomic Research Centre,  \\
Trombay, Mumbai 400085, India \\
Homi Bhabha National Institute, Anushakti Nagar,
Mumbai 400094, India }
\end {center}

\begin {abstract}
The cross section of the photonuclear reaction in the nucleon-resonance region is calculated to search the 
resonance-nucleon scattering amplitude in the nucleus. It is assumed that the resonance $R$ is produced and 
decayed in the elementary photon-nucleon $(\gamma N)$ reaction in the nucleus as $\gamma N \to R \to \gamma N$. 
The resonance interacts with the nucleons while propagating through the nucleus. Therefore, the calculated 
photonuclear reaction cross section is compared with the data to extract the resonance-nucleon scattering 
parameters in the nucleus.
\end {abstract}



\section{Introduction}

The photonuclear reaction is a pertinent probe to investigate various aspects of basic research and application 
in physics, see Ref.~\cite{Zilges} for review. This reaction is a key ingredient to investigate the nucleosynthesis
is in the universe. The photonuclear reaction triggered by the lighting discharge \cite{Enoto} has been used to 
observe the previously unknown channels for generating carbon, nitrogen and oxygen isotopes. 
Since 
the photon induced nuclear reaction does not create distortion in the entrance channel, it is a potential tool 
to study the final state interactions of the produced hadrons \cite{Tarasov}. The nuclear transparency measured 
for the photoproduced $\omega$ and $\phi$ mesons illustrates the $\omega N$ and $\phi N$ cross sections in the 
nucleus are larger than the respective cross sections in the free space \cite{DasA}. The photonuclear reaction 
is in extensive use to explore the color transparency of hadrons \cite{Dutta} (also see the references therein).

The remarkable observation in the measured energy distribution spectrum of the photonuclear reaction in the 
resonance region ($\approx 0.2-1.2$ GeV) \cite{Bianchi, Muccifora} is the disappearance of certain resonances 
which are distinctly visible in that measured for the photonucleon reaction \cite{Zyla, Armstrongp, Armstrongn}. 
The spectra measured for nuclei show the $\Delta(1232)$-peak is partially modified whereas the peaks of higher 
resonances (e.g., $N(1520)$, .....) are vanished in the nucleus. 
The 
change in the spectral-shape of resonances in a nucleus occurs because of the nuclear medium effect on them. 
The Fermi-motion of the nucleon in a nucleus produces the suppression and broadening of the resonance peak. 
The resonance-nucleus interaction can also dampen and widen the peak of the resonance, which is addressed as 
the collision broadening of the resonance. It linearly depends on the in-medium resonance-nucleon cross section, 
i.e., $\Gamma^c_R \propto \sigma_t^{*RN}$ \cite{SDasR}.

The photonuclear reaction in the resonance region has been studied by Rapp et al., \cite{Rapp} using the 
vector-meson dominance model which states the photon is converted to the transverse $\rho$-meson. Hence, the 
$\gamma$-nucleon resonances have been interpreted as the $\rho$-nucleon resonances. The parameters appearing 
in this calculation are extracted by fitting the calculated photoproton reaction cross section with the data. 
Using 
those parameters, the cross section of the photonuclear reaction is evaluated. The Fermi-motion of the 
nucleon in the nucleus is not considered in the calculation. The nuclear effect is evaluated by the Lindhard
function describing the resonance dynamics in the nucleus. The calculated results are shown in well accord with 
the data reported for nuclei \cite{Rapp}.
Kondratyuk 
et al., \cite{Kondratyuk} have also calculated the cross section of the photonuclear reaction in the quoted 
energy region by evaluating the forward Compton scattering amplitude of the reaction. They have extracted the 
$\gamma$-nucleon resonances' parameters (i.e., mass and width) by fitting the calculated photonucleon reaction 
cross sections with the data, and used those to evaluate the cross section of the photonuclear reaction. 
The 
effect of the Fermi-motion of the nucleon is estimated by the Fermi-suppression factor. The resonance-nucleus 
interaction is described by the optical potential generated by folding the resonance-nucleon scattering 
amplitude with the nuclear density. The resonance-nucleon cross section in the nucleus is extracted from the 
measured cross section of the photonuclear reaction.

The cross sections of the photonuclear reaction is also calculated in this paper. The Fermi-suppression factor 
evaluated by Kondratyuk et al., \cite{Kondratyuk}, as mentioned above,  is used to assess the effect of the 
Fermi-motion of the bound nucleon. The optical potential for the resonance-nucleus interaction is also formulated 
by folding the resonance-nucleon scattering amplitude with the nuclear density. 
The 
forward Compton scattering amplitude of both the photonucleon and photonuclear reactions (used to calculate the 
cross section) differs from that done by Kondratyuk et al., \cite{Kondratyuk} in the following ways. The measured 
mass and width of all four-starred resonances \cite{Zyla} (except those of the $\Delta(1232)$ resonance) are used 
in the calculation. 
The 
photon-nucleon interactions are described by the Lagrangians, which are used to evaluate the spin-isospin matrix 
elements associated with the production and decay of the resonances. The calculated cross sections of the 
photonucleon reactions are fitted with the data to extract the resonance-nucleon-photon coupling-constants  
appearing in the Lagrangians.
The 
propagator of the resonance is taken as the eikonal form in the Glauber model \cite{SDasR}, which illustrates 
the propagation of the resonance from the production point to the decay point in the nucleus. The calculated 
results of the photonuclear reaction are fitted with the measured cross sections \cite{Bianchi} to extract the 
resonance-nucleon scattering amplitude (not only the cross section) in the nucleus.

\section{Formalism}

The Lagrangian describing the resonance $R$ production due to the nucleon-photon $N\gamma$ coupling is categorized 
based on the positive or negative parity state of $R$. For the positive parity state, the Lagrangian describing 
the $RN\gamma$ coupling is given by \cite{Rapp, Peters}
\begin{equation}
{\cal L}_{RN\gamma}
=\frac{f_{RN\gamma}}{m_\pi}
          R^\dagger ( {\bf s^\dagger} \times {\bf k}_\gamma )
 \cdot {\bf \epsilon} ({\bf \lambda}_\gamma) i_r^\dagger N.
\label{LRp}                                            
\end{equation}
The spin operator 
${\bf s^\dagger} = {\bf \sigma} ({\bf S^\dagger})$ connects 
$N$ of $J^P = \frac{1}{2}^+$ to $R$ of
$J^P = \frac{1}{2}^+ (\frac{3}{2}^+)$, where $J^P$ denotes spin$^{parity}$ of the particle.
$i_r^\dagger = \tau_r (I_r^\dagger)$ does the isospin $\frac{1}{2} \to \frac{1}{2} (\frac{3}{2})$ transition of 
the particle.

For $R$ of $J^P=\frac{5}{2}^+$, the Lagrangian is expressed by Friman et al., \cite{Friman} as
\begin{equation}
{\cal L}_{RN\gamma} = \frac{f_{RN\gamma}}{m_\pi} 
 R^\dagger S^\dagger_{ij} k_{\gamma ,i} \epsilon_j i_r^\dagger N,
\label{LRpp}                                            
\end{equation}
where $S^\dagger_{ij}$ denotes the spin $\frac{1}{2}^+ \to \frac{5}{2}^+$ transition operator. It is a tensor 
of rank 2 and $6 \times 2$ matrix in spin space. The spin transition matrix \cite{Friman} is 
\begin{equation}
<R| S^\dagger_{ij} k_{\gamma ,i} \epsilon_j |N> 
= <X_{ij}| k_{\gamma ,i} \epsilon_j |N>.
\label{MRpp}                                            
\end{equation}
$|X_{ij}>$ can be constructed by coupling of two spin-1 objects $\epsilon^s_i $ to the nucleon spinor 
$\chi^{m_s}_{1/2}$ \cite{Friman}, i.e., 
\begin{equation}
|X^M_{ij}> = \sum_{r,s,t,m_s} C(2r\frac{1}{2}m_s|\frac{5}{2}M) C(1s1t|2r) 
\epsilon^s_i \epsilon^t_j |\chi^{m_s}_{1/2}>,
\label{XRpp}                                            
\end{equation}
where the unit vectors are given by
\begin{equation}
\epsilon^{+1}=-\frac{1}{2}(1,i,0), ~~ \epsilon^0=(0,0,1), ~~
\mbox{and} ~~ \epsilon^{-1}=\frac{1}{2}(1,-i,0).
\label{eRpp}                                            
\end{equation}
The evaluation of the matrix element for spin-$\frac{5}{2}$ object has been discussed elaborately in 
Ref.~\cite{Friman}.

The Lagrangian describing the transition $N$ of $J^P = \frac{1}{2}^+$ to $R$ of 
$J^P = \frac{1}{2}^- (\frac{3}{2}^-)$ is given by \cite{Rapp, Peters}
\begin{equation}
{\cal L}_{RN\gamma}
=\frac{f_{RN\gamma}}{m_\pi}  R^\dagger
 ( E_\gamma {\bf s^\dagger} \cdot {\bf \epsilon} ({\bf \lambda}_\gamma) - \epsilon^0 ({\bf \lambda}_\gamma) 
{\bf s^\dagger} \cdot {\bf k}_\gamma ) i_r^\dagger N,
\label{LRs}                                            
\end{equation}
where $\epsilon^0$ (the time-component of photon polarization) is zero for the real photon, as it has only 
transverse polarizations \cite{Halzen}. The coupling constants $f_{R\gamma N}$ appearing in the Lagrangians 
${\cal L}_{RN\gamma}$ are discussed later.

The nucleus $A$ is assumed as a composition of the nucleon $N$ and nucleus $B$. The photon $\gamma$ couples to 
$N$ and produces the resonance $R$ which (after propagating certain distance) decays to $N^\prime \gamma^\prime$.  
The nucleon $N^\prime$ recombines with $B$ to form the nucleus $A^\prime$ in the final state. Symbolically, it 
can be represented as $\gamma A \to RB  \to \gamma^\prime A^\prime$. The $T$-matrix of the reaction can be 
written as
\begin{equation}
T^{\gamma A}_{fi}
= \int \int d{\bf r^\prime} d{\bf r} \sum_{R,B} 
 <\gamma^\prime A^\prime |\Gamma^\dagger_{RN\gamma}| RB>
   G_R({\bf r}^\prime -{\bf r}) 
  <BR |\Gamma_{RN\gamma}| \gamma A>,
\label{TfiA}
\end{equation}
where $\Gamma_{RN\gamma}$ denotes the vertex operator. It can be expressed by the Lagrangians in Eqs.~(\ref{LRp}), 
(\ref{LRpp}) and (\ref{LRs}) without the wave-functions of the nucleon and resonance included in it.

$G_R ({\bf r^\prime}-{\bf r})$ in Eq.~(\ref{TfiA}) represents the propagation of the resonance $R$ from the 
production (space) coordinate ${\bf r}$ to the decay coordinate ${\bf r^\prime}$. Since the forward $T$-matrix is 
related the cross section of the reaction, the eikonal approximation in the Glauber model is used to expressed 
$G_R ({\bf r^\prime}-{\bf r})$ \cite{SDasR}:
\begin{equation}
G_R({\bf r}^\prime -{\bf r})
= \delta ({\bf b^\prime} -{\bf b}) \theta (z^\prime -z)
   e^{ i{\bf k}_R.({\bf r^\prime} - {\bf r}) } 
   D_{{\bf k}_R} ({\bf b},z^\prime,z),
\label{Gres}
\end{equation}
where ${\bf b} ({\bf b^\prime})$ is the impact factor for ${\bf r} (\bf r^\prime)$ in the cylindrical coordinate 
system. The non-relativistic form of $D_{{\bf k}_R} ({\bf b}, z^\prime,z)$ is given by
\begin{equation}
D_{{\bf k}_R} ({\bf b},z^\prime,z)
= -\frac{i}{v_{R\parallel}} 
    exp \left [ \frac{i}{v_{R\parallel}} 
          \int^{z^\prime}_z dz^{\prime \prime} 
    \{ \Delta(m) - V_{OR} ({\bf b}, z^{\prime \prime}) \}    
   \right ]  
\label{Dres}
\end{equation}
with $\Delta(m) = m-m_R+\frac{i}{2}\Gamma_R$. $m$ is the invariant mass of the decay-products (i.e., $N\gamma$ 
in the considered reaction) of the resonance $R$. $m_R$ and $\Gamma_R$ are the resonant mass and width of $R$, 
as listed in table-\ref{tbR}. $V_{OR}$ denotes the resonance-nucleus optical potential elaborated later. 
$v_{R\parallel}$ is the velocity of the resonance ($v_R$) parallel to the beam direction.

\begin{table}[h]
\caption{ Four-starred resonances $R$ taken from Ref.~\cite{Zyla}. $J^P$ denotes $spin^{parity}$ of $R$. $m_R$ 
and $\Gamma_R$ are the mass and width respectively of $R$ used in the calculation. The range of the measured 
$m_R$ and $\Gamma_R$ are given in the bracket. }
\begin{center}
\begin{tabular} {|c|c|c|c|}
\hline
 $J^P$           & $R$            & $m_R$(MeV)       & $\Gamma_R$(MeV) \\
\hline                                                       
 $~$             & $N(1440)$      & 1440 (1410-1470) & 350 (250-450)  \\
 $\frac{1}{2}^+$ & $N(1710)$      & 1710 (1680-1740) & 140 (80-200)   \\         
 $~$             & $\Delta(1910)$ & 1910 (1850-1950) & 300 (200-400)  \\
\hline
 $~$             & $\Delta(1232)$ & 1220 (1230-1234) & 110 (114-120)  \\
 $\frac{3}{2}^+$ & $\Delta(1600)$ & 1600 (1500-1640) & 250 (200-300)  \\
 $~$             & $N(1720)$      & 1720 (1680-1750) & 250 (150-400)  \\
 $~$             & $N(1900)$      & 1900 (1890-1950) & 200 (100-320)  \\
\hline
 $\frac{5}{2}^+$ & $N(1680)$      & 1680 (1680-1690) & 120 (115-130)  \\
 $~$             & $\Delta(1905)$ & 1905 (1855-1910) & 330 (270-400)  \\
\hline
\hline                                                        
 $~$             & $N(1535)$      & 1535 (1515-1545) & 150 (125-175)  \\
 $\frac{1}{2}^-$ & $\Delta(1620)$ & 1620 (1590-1630) & 130 (110-150)  \\
 $~$             & $N(1650)$      & 1650 (1635-1665) & 125 (100-150)  \\
 $~$             & $N(1895)$      & 1895 (1870-1920) & 120 (80-200)   \\
\hline
$\frac{3}{2}^-$  & $N(1520)$      & 1510 (1510-1520) & 110 (100-120)  \\
$~$              & $\Delta(1700)$ & 1700 (1690-1730) & 300 (220-380)  \\
\hline
\end{tabular}
\end{center}
\label{tbR}
\end{table}

After taking the average over the initial state and the summation over the final state, the $T$-matrix for the 
elastic photon scattering on a spin zero nucleus can be written as
${\bar T}^{\gamma A}_{ii} = \frac{1}{2} \sum_{\lambda_\gamma \lambda_{\gamma^\prime}} T^{\gamma A}_{ii}$, 
where 
$\lambda_{\gamma (\gamma^\prime)}$ denotes the polarization of the incoming (outgoing) photon.  The Compton 
scattering amplitude $F^{\gamma A}_{ii}$ is related to ${\bar T}^{\gamma A}_{ii}$ as 
$F^{\gamma A}_{ii} = \frac{m_A}{4\pi\sqrt{s}} {\bar T}^{\gamma A}_{ii}$ \cite{Ericson}, where $m_A$ is the mass 
of the target and $\sqrt{s}$ denotes the total energy in the center of mass system. In the considered energy 
region (i.e., $\leq 1.2$ GeV), the above equation can be reduced to 
$F^{\gamma A}_{ii} \approx \frac{1}{4\pi} {\bar T}^{\gamma A}_{ii}$ for a nucleus \cite{Ericson}. The total 
scattering cross section $\sigma^{\gamma A}_t (R)$ of the photonuclear reaction due to the resonance $R$ is 
\begin{equation}
\sigma_t^{\gamma A} (R)
= \frac{4\pi}{k_\gamma} Im [F^{\gamma A}_{ii}].
\label{XTgA}
\end{equation}

The above formalism can be modified to calculate the cross section $\sigma_t^{\gamma N}$ of the photonucleon 
reaction (i.e., $\gamma N \to R \to \gamma^\prime N^\prime$), where the Fermi-motion of the nucleon and the 
resonance-nucleus potential do not exist. The $T$-matrix for this reaction can be written as
\begin{equation}
T^{\gamma N}_{ii}
= \int \int d{\bf r^\prime} d{\bf r} \sum_R 
 <\gamma^\prime N^\prime |\Gamma^\dagger_{RN\gamma}| R>
   G_R({\bf r}^\prime -{\bf r}) 
  <R |\Gamma_{RN\gamma}| \gamma N>.
\label{TfiN}
\end{equation}
$V_{OR}$ appearing in $G_R({\bf r}^\prime -{\bf r})$, see Eqs.~(\ref{Dres}) and (\ref{Gres}), is equal to zero 
for the photonucleon reaction. To evaluate $\sigma_t^{\gamma N}$, the polarizations of the photons and the spins 
of the nucleons in the initial and final states have to be accounted for taking the average over the initial 
state and the summation over the final state.

\section{Results and Discussion}

The cross sections of the photonucleon and photonuclear reactions have been calculated in the resonance region, 
i.e., $E_\gamma \approx 0.2-1.2$ GeV. The background contributions to the cross sections are taken from 
Ref.~\cite{Kondratyuk}. The resonance-nucleon-photon coupling-constants $f_{RN\gamma}$ in the calculated cross 
sections of the photonucleon reactions are adjusted so that the calculated results added with the background of 
the reactions reproduce the data. Those values of $f_{RN\gamma}$ are used to evaluate the cross sections of the 
photonuclear reactions.

\begin{figure}[h]
\begin{center}
\centerline {\vbox {
\psfig{figure=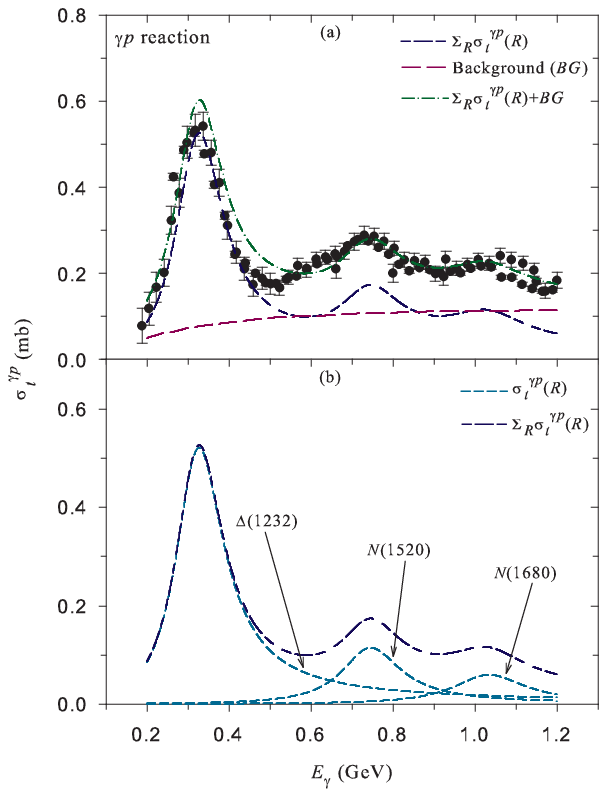,height=14.0 cm, width=12.0 cm}
}}
\caption{
(color online).
Upper part (a): 
The calculated total cross section $\sigma_t^{\gamma p}$ of the photoproton $\gamma p$ reaction. The 
short-long-short-dashed curve occurs due to the contributions of all resonances, as listed in 
table~\ref{tbR}, to the calculated cross section. 
The 
long-dashed curve represents the background contribution to the cross section \cite{Kondratyuk}. The 
dot-dashed curve arises due to the addition of the above curves. The data are taken from 
Refs.~\cite{Zyla, Armstrongp}.
Lower part (b): 
The short-dashed curves represent the cross sections due to the $\Delta (1232)$, $N(1520)$ and $N(1680)$ 
resonances. The short-long-short curve is already described in the upper part (a) of the figure.  
}
\label{gproton}
\end{center}
\end{figure}

The calculated total cross sections of the photoproton $\gamma p$ reaction are presented in Fig.~\ref{gproton}, 
where $\sigma^{RN}_t (R)$ represents the cross section of the $\gamma p$ reaction occurring because of the 
individual resonance $R$. The short-long-short-dashed curve [labeled as $\Sigma_R \sigma^{RN}_t (R)$] in 
Fig.~\ref{gproton}(a) denotes the calculated cross section due to all resonances, as listed in table-\ref{tbR}. 
The long-dashed curve (labeled as $BG$) refers to the background of the reaction \cite{Kondratyuk}.
The 
dot-dashed curve (labeled as $\Sigma_R \sigma^{RN}_t (R) + BG$) in Fig.~\ref{gproton}(a) arising due to the 
addition of the calculated results (short-long-short-dashed curve) and the background (long-dashed curve) is 
compared with the measured cross section of the $\gamma p$ reaction \cite{Zyla, Armstrongp}. The extracted values 
of the resonance-proton-photon coupling-constants $f_{Rp\gamma}$ are listed in table-\ref{tbC}.
The 
short-long-short-dashed curve in Fig.~\ref{gproton}(b) is already explained in Fig.~\ref{gproton}(a). The 
dominant contributions to the cross section of the $\gamma p$ reaction, as represented by the short-dashed 
curves [labeled as $\sigma^{RN}_t (R)$] in Fig.~\ref{gproton}(b), arise due to the $\Delta(1232)$, $N(1520)$, 
and $N(1680)$ resonances. Amongst them, the cross section because of the $\Delta(1232)$ resonance is distinctly 
largest. The cross sections of the above reaction due to other resonances are not shown explicitly, as their 
contributions to the cross section are less significant.

\begin{figure}[h]
\begin{center}
\centerline {\vbox {
\psfig{figure=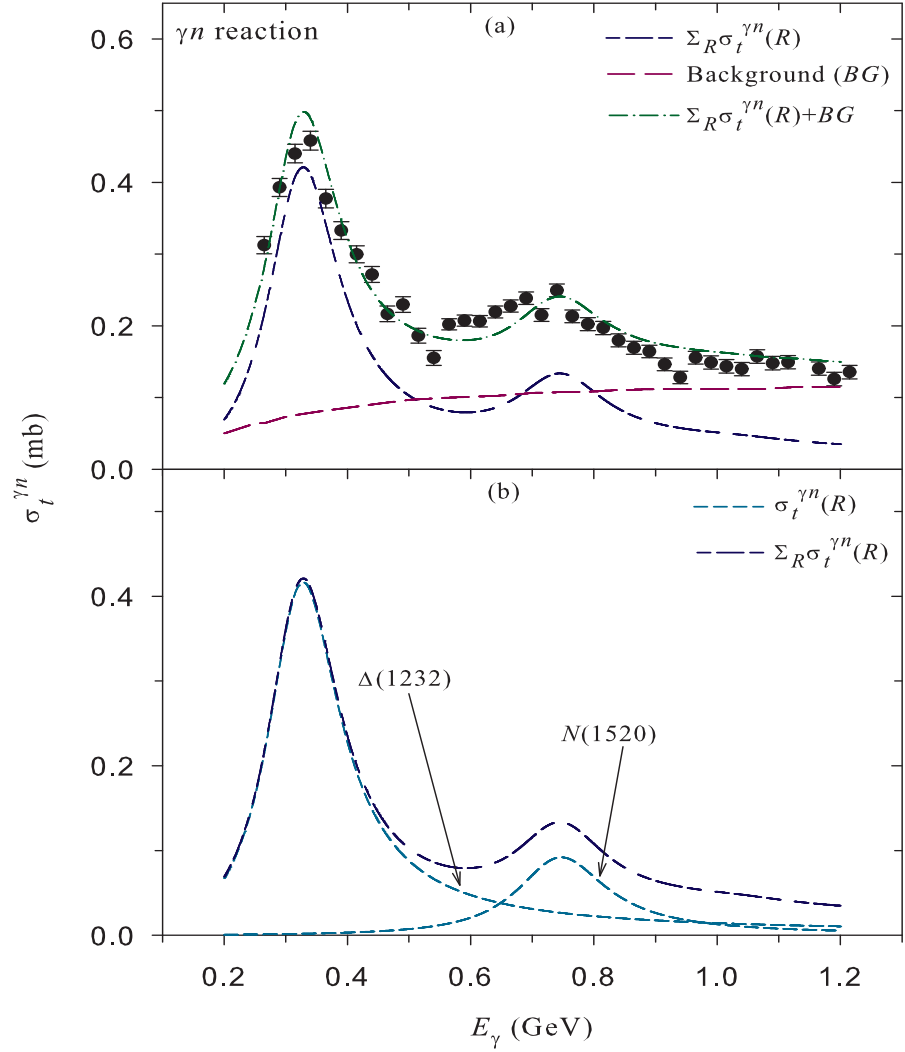,height=14.0 cm, width=12.0 cm}
}}
\caption{
(color online).
As in Fig.~\ref{gproton} but for the photoneutron $\gamma n$ reaction, see text. The data are taken from 
Ref.~\cite{Armstrongn}.
}
\label{gneutron}
\end{center}
\end{figure}

The total cross sections of the photoneutron $\gamma n$ reaction have been shown in Fig.~\ref{gneutron}. The 
curves appearing in this figure represent the $\gamma n$ reaction cross sections similar to those explained in 
Fig.~\ref{gproton} for the $\gamma p$ reaction, except the cross section due to the $N(1680)$ resonance in the 
$\gamma n$ reaction is insignificant [could not be shown Fig.~\ref{gneutron}(b)] in contrast to that occurring 
in the $\gamma p$ reaction [see Fig.~\ref{gproton}(b)]. The resonance-neutron-photon coupling-constants 
$f_{Rn\gamma}$ in the calculated cross section of the $\gamma n$ reaction, as shown in Fig.~\ref{gneutron}(a), 
have been determined using the data taken from Ref.~\cite{Armstrongn}. The values of $f_{Rn\gamma}$ are also 
listed in table-\ref{tbC}.

\begin{table}[h]
\caption{ The resonance-nucleon-photon coupling constants $f_{RN\gamma}$ appearing in the Lagrangians in 
Eqs.~(\ref{LRp}), (\ref{LRpp}) and (\ref{LRs}), as extracted from the measured cross sections of the photonucleon 
reactions. }
\begin{center}
\begin{tabular} {|c|c|c|}
\hline
 $~$             & $ f_{p(1440)p\gamma} = 1.28\times 10^{-2} $            & 
                   $ f_{n(1440)n\gamma} = 1.68\times 10^{-2} $           \\              
 $\frac{1}{2}^+$ & $ f_{p(1710)p\gamma} = 1.22\times 10^{-3} $            & 
                   $ f_{n(1710)n\gamma} = 1.22\times 10^{-3} $           \\         
 $~$             & $ f_{\Delta^+(1910)p\gamma} = 3.85\times 10^{-3} $     & 
                   $ f_{\Delta^0(1910)n\gamma} = 3.85\times 10^{-3} $    \\

\hline
 $~$             & $ f_{\Delta^+(1232)p\gamma} = 0.114 $                  & 
                   $ f_{\Delta^0(1232)n\gamma} = 0.102 $                 \\
 $\frac{3}{2}^+$ & $ f_{\Delta^+(1600)\gamma p} = 2.86\times 10^{-3} $    &       
                   $ f_{\Delta^0(1600)\gamma n} = 2.86\times 10^{-3} $   \\
 $~$             & $ f_{p(1720)\gamma p} = 6.21\times 10^{-3} $           & 
                   $ f_{n(1720)\gamma n} = 3.92\times 10^{-3} $          \\  
 $~$             & $ f_{p(1900)\gamma p} = 1.42\times 10^{-3} $           & 
                   $ f_{n(1900)\gamma n} = 1.8 \times 10^{-3} $          \\  
\hline
 $\frac{5}{2}^+$ & $ f_{p(1680)\gamma p} = 3.21\times 10^{-2} $           & 
                   $ f_{n(1680)\gamma n} = 6.42\times 10^{-3} $          \\  
 $~$             & $ f_{\Delta^+(1905)\gamma p} = 2.44\times 10^{-2} $    &    
                   $ f_{\Delta^0(1905)\gamma n} = 1.57\times 10^{-2} $   \\
\hline
\hline
 $~$             & $ f_{p(1535)\gamma p} = 8.27\times 10^{-3} $           & 
                   $ f_{n(1535)\gamma n} = 3.7 \times 10^{-3} $          \\              
 $\frac{1}{2}^-$ & $ f_{\Delta^+(1620)p\gamma} = 6.33\times 10^{-3} $     & 
                   $ f_{\Delta^0(1620)n\gamma} = 6.33\times 10^{-3} $     \\         
 $~$             & $ f_{p(1650)p\gamma} = 5.59\times 10^{-3} $            & 
                   $ f_{n(1650)n\gamma} = 1.25\times 10^{-3} $           \\
 $~$             & $ f_{p(1895)p\gamma} = 2.02\times 10^{-3} $            & 
                   $ f_{n(1895)n\gamma} = 1.11\times 10^{-3} $           \\
\hline
 $\frac{3}{2}^-$ & $ f_{p(1520)p\gamma} = 3.2 \times 10^{-2} $            &       
                   $ f_{n(1520)n\gamma} = 2.57\times 10^{-2} $           \\
 $~$             & $ f_{\Delta^+(1700)p\gamma} = 1.48\times 10^{-2} $     & 
                   $ f_{\Delta^0(1700)n\gamma} = 6.05\times 10^{-3} $    \\  
 \hline
\end{tabular}
\end{center}
\label{tbC}
\end{table}

The cross sections of the photonuclear reactions have been calculated for nuclei, where the nuclear effect, 
i.e., the Fermi-motion of the bound nucleon and the resonance-nucleus interaction, are investigated. The 
Fermi-motion causes the suppression and broadening of the cross section. To describe it, the Fermi-suppression 
factor estimated by Kondratyuk et al., (as shown in Fig.~3 in Ref.~\cite{Kondratyuk}) has been incorporated in 
the calculation. 
The 
interaction of the resonance with the nucleus is addressed by the resonance-nucleus optical potential $V_{OR}$, 
which is generated by folding the forward resonance-nucleon scattering amplitude $f_{RN}$ (in the free space) 
with the density of nucleus \cite{Glauber}, i.e.,
\begin{equation}
 -\frac{1}{v_R} V_{OR} ({\bf r})
= \frac{2\pi}{k_R} f_{RN}(0) \varrho ({\bf r}) 
= \frac{1}{2} [\alpha_{RN} + i] 
  \sigma_t^{RN} \varrho ({\bf r}).
\label{VOR}
\end{equation}
$\varrho ({\bf r})$ denotes the spatial distribution of the nuclear density, as extracted from the electron 
scattering data \cite{Jager}. In fact, the electron scattering determines the proton density of the nucleus. 
Therefore, $\varrho ({\bf r})$ is approximated by the proton density with the normalization: 
$ 4\pi \int r^2 \varrho (r) dr = A $ (mass number of the nucleus). 
$\alpha_{RN}$ 
represents the ratio of the real to imaginary parts of $f_{RN}$. $\sigma_t^{RN}$ is the total resonance-nucleon 
scattering cross section: $\sigma_t^{RN} = \frac{4\pi}{k_R} Im [f_{RN}]$. The above form of the optical potential 
of the hadron is largely used to study the nuclear reactions \cite{Frankfurt}.
These quantities in the nucleus (symbolized by $\alpha_{RN}^*$ and  $\sigma_t^{*RN}$), as done later, have been 
extracted by fitting the calculated photonuclear cross section with that of the measured value. $V_{OR}$ with 
$\alpha_{RN}^*$ and $\sigma_t^{*RN}$ defines the resonance-nucleus potential in the reaction.  
Since
$V_{OR}$ is absorptive, the peak cross section of the reaction due to the resonance $R$ is attenuated by the 
factor $\frac{\sigma_t^{\gamma A} (R)_{V_{OR}\neq 0}}{\sigma_t^{\gamma A} (R)_{V_{OR}= 0}}$ because of the
inclusion of the potential in the calculation.

\begin{figure}[h]
\begin{center}
\centerline {\vbox {
\psfig{figure=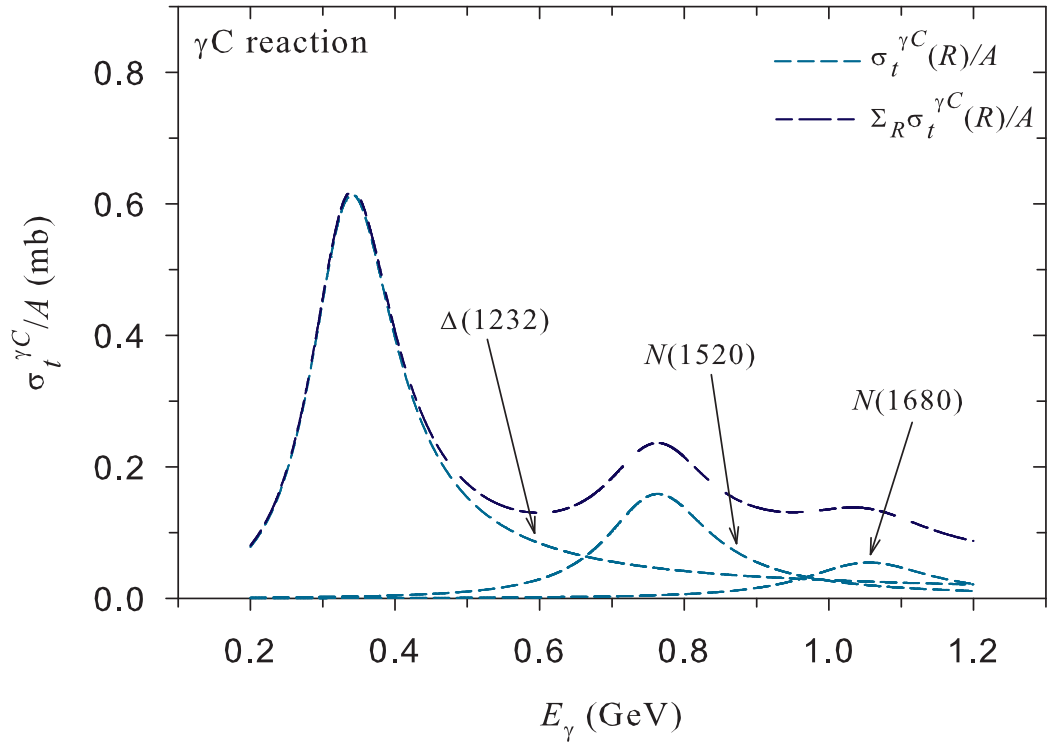,height=10.0 cm, width=14.0 cm}
}}
\caption{
(color online).
Total cross section per nucleon $\sigma_t^{\gamma C}/A$ of the photocarbon $\gamma$C reaction calculated 
without considering the Fermi-motion and the resonance-nucleus interaction. The curves appearing in the 
figure qualitatively represent those explained in Fig.~\ref{gproton}(b).
}
\label{XTC00}
\end{center}
\end{figure}

The cross section per nucleon $\sigma_t^{\gamma C}/A$ of the photocarbon $\gamma$C reaction is calculated without 
considering the Fermi-motion of the bound nucleon and the resonance-nucleus interaction [i.e., potential $V_{OR}$ 
in Eq.~(\ref{VOR})]. The resonance-nucleon-photon coupling-constants $f_{RN\gamma}$ extracted from the 
photonucleon reactions, as listed in table-\ref{tbC}, are used in the calculation.
The 
calculated results versus the beam energy $E_\gamma$ are presented in Fig.~\ref{XTC00}. The short-dashed and the 
short-long-short curves appearing in the figure qualitatively illustrate those explained in Fig.~\ref{gproton}(b) 
for the $\gamma p$ reaction, i.e., the dominant contributions to the cross section of the $\gamma C$ reaction 
arise due to the $\Delta(1232)$, $N(1520)$ and $N(1680)$ resonances.

\begin{figure}[h]
\begin{center}
\centerline {\vbox {
\psfig{figure=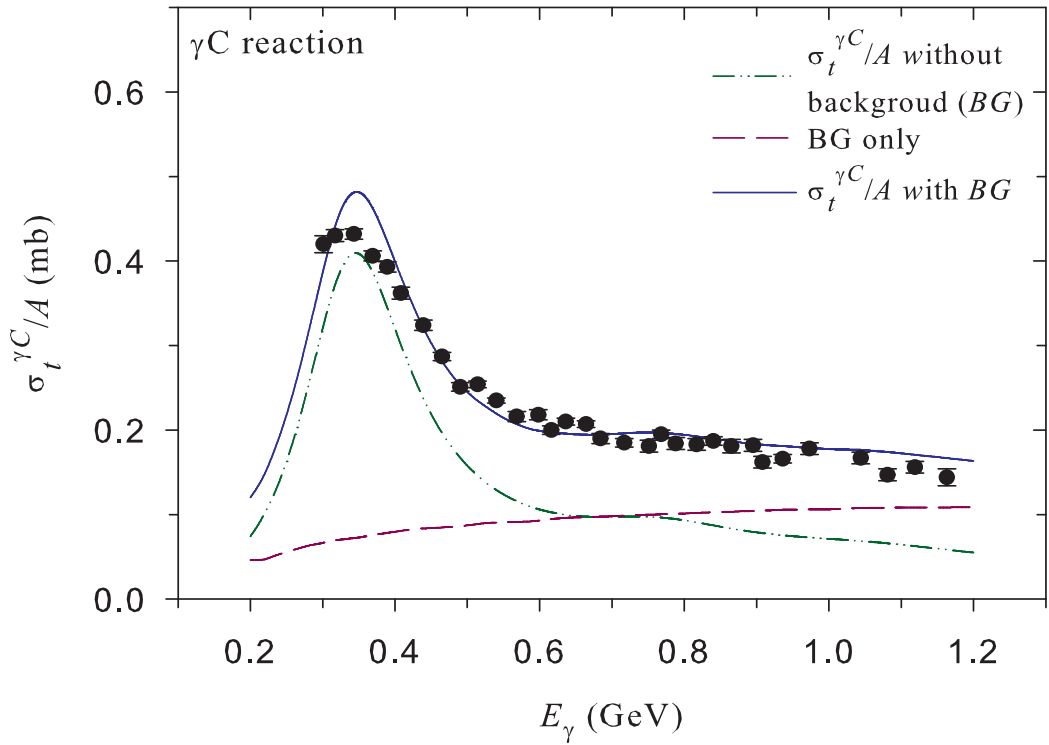,height=10.0 cm, width=14.0 cm}
}}
\caption{
(color online).
The cross section/nucleon $\sigma_t^{\gamma C}/A$ due to all resonances calculated using the Fermi-motion of 
the bound nucleon, and the resonance-nucleus interaction potential is represented by the dot-dot-dashed curve.
The
long-dashed curve is the background contribution to the cross section \cite{Kondratyuk}. The solid curve 
arises because of the summation of the above curves. See text for the detail. The data are taken from 
Ref.~\cite {Bianchi}.
}
\label{XTgC}
\end{center}
\end{figure}

The photonuclear cross section $\sigma_t^{\gamma C}/A$ due to all resonances for the carbon nucleus is evaluated 
including the Fermi-motion of the bound nucleon and the resonance-nucleus optical potential in the calculation. 
The Fermi-motion, as mentioned earlier, is addressed by the Fermi-suppression factor \cite{Kondratyuk}. 
In
Fig.~\ref{XTgC}, the calculated cross sections $\sigma_t^{\gamma C}/A$ is represented by the dot-dot-dashed 
curve and the background of the reaction (as given in Ref.~\cite{Kondratyuk}) is denoted by the long-dashed 
curve. The calculated results added with the background, as represented by the solid curve, are fitted with 
the data \cite{Bianchi} to extract the resonance-nucleon scattering parameters [i.e., $\alpha_{RN}^*$ and 
$\sigma_t^{*RN}$ as described in Eq.~(\ref{VOR})] in the nucleus. 
As 
shown in Fig.~\ref{XTC00}, the $\Delta(1232)$, $N(1520)$ and $N(1680)$ resonances contribute dominantly to the 
cross section. Therefore, $\alpha_{RN}^*$ and $\sigma_t^{*RN}$ for these resonances are quoted in 
table-\ref{tbsp}.

\begin{table}[h]
\caption{ The scattering parameters, i.e., $\alpha_{RN}^*$ and $\sigma_t^{*RN}$, of the $\Delta (1232)$,  
N(1520) and N(1680) resonances in the nucleus. $\sigma_t^{*RN}$ estimated by Kondratyuk et al., 
\cite{Kondratyuk} is also compared. }
\begin{center}
\begin{tabular} {|c|c|c|c|}
\hline
  $R$              & $\alpha_{RN}^*$  & $\sigma_t^{*RN}$ (mb) & $\sigma_t^{*RN}$ (mb) in Ref.~\cite{Kondratyuk} \\ 
\hline                                                       
  $\Delta(1232)$   &     1.34         &   54.11               &     78.94                                  \\
  $N(1520)$        &     0.57         &   426.25              &    183.36                                  \\         
  $N(1680)$        &    -0.31         &   425.33              &    153.06                                  \\ 
\hline
\end{tabular}
\end{center}
\label{tbsp}
\end{table}

\begin{figure}[h]
\begin{center}
\centerline {\vbox {
\psfig{figure=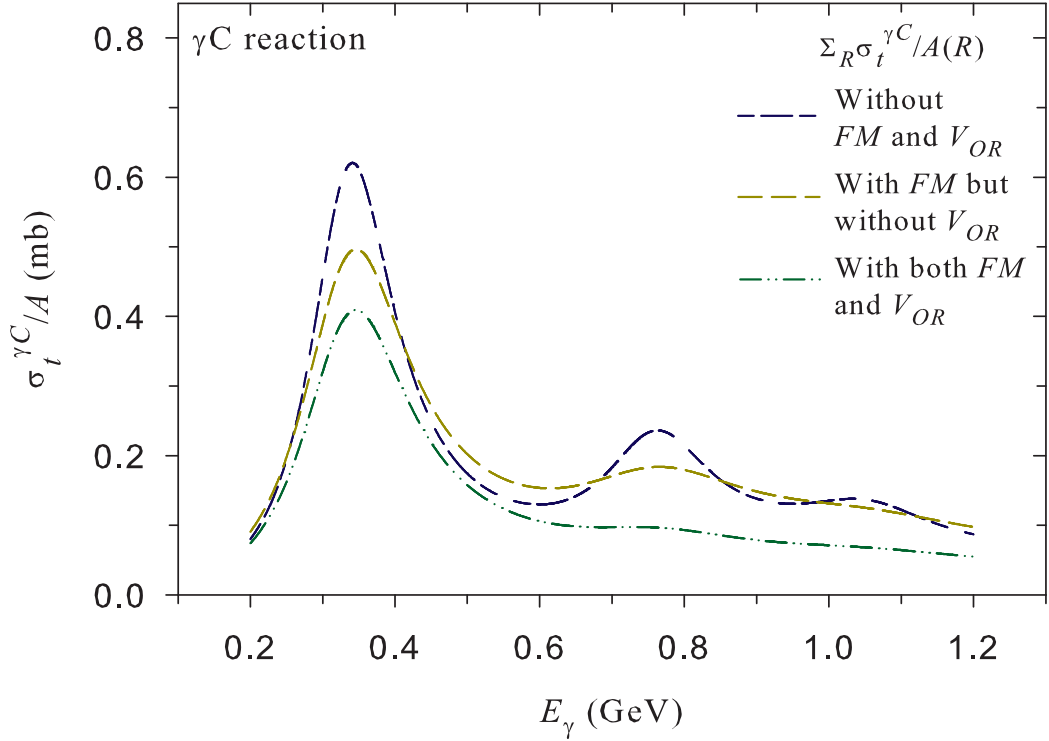,height=10.0 cm, width=14.0 cm}
}}
\caption{
(color online).
The short-long-short-dashed curve represents the calculated total cross section per nucleon 
$\sigma_t^{\gamma C}/A$ due to all resonances, as done in Fig.~\ref{XTC00}. The medium-dashed curve denotes 
$\sigma_t^{\gamma C}/A$ evaluated incorporating the Fermi-motion ($FM$) of the bound nucleon, where as the 
dot-dot-dashed curve (also shown in Fig.~\ref{XTgC}) represents that assessed including both $FM$ and the 
resonance-nucleus optical potential $V_{OR}$ in the calculation.
}
\label{XTCFP}
\end{center}
\end{figure}

The sensitivities of the Fermi-motion ($FM$) of the bound nucleon and the resonance-nucleus potential ($V_{OR}$) 
to the calculated cross section $\sigma_t^{\gamma C}/A$ (without the background contribution) are exhibited in 
Fig.~\ref{XTCFP}. $V_{OR}$ is estimated using the extracted values of $\alpha_{RN}^*$ and $\sigma_t^{*RN}$ for 
the resonances. The short-long-short-dashed curve (also shown in Fig.~\ref{XTC00}) represents 
$\sigma_t^{\gamma C}/A$ of the $\gamma$C reaction, where $FM$ and $V_{OR}$ are not considered in the calculation. 
The 
medium-dashed curve results because of the inclusion of the Fermi-suppression factor due to the Fermi-motion 
\cite{Kondratyuk}, as described earlier, in the calculated cross section. The suppression and broadening of the 
resonance peaks due to the Fermi-motion is distinctly visible in Fig.~\ref{XTCFP}. In fact, the peaks in the 
region of $N(1520)$ and $\Delta(1700)$ resonances are smeared out significantly due to the Fermi-motion. 
The 
dot-dot-dashed curve (also appeared in Fig.~\ref{XTgC}) denotes the calculated $\sigma_t^{\gamma C}/A$ due to 
both $FM$ and $V_{OR}$ incorporated in the calculation.

\begin{figure}[h]
\begin{center}
\centerline {\vbox {
\psfig{figure=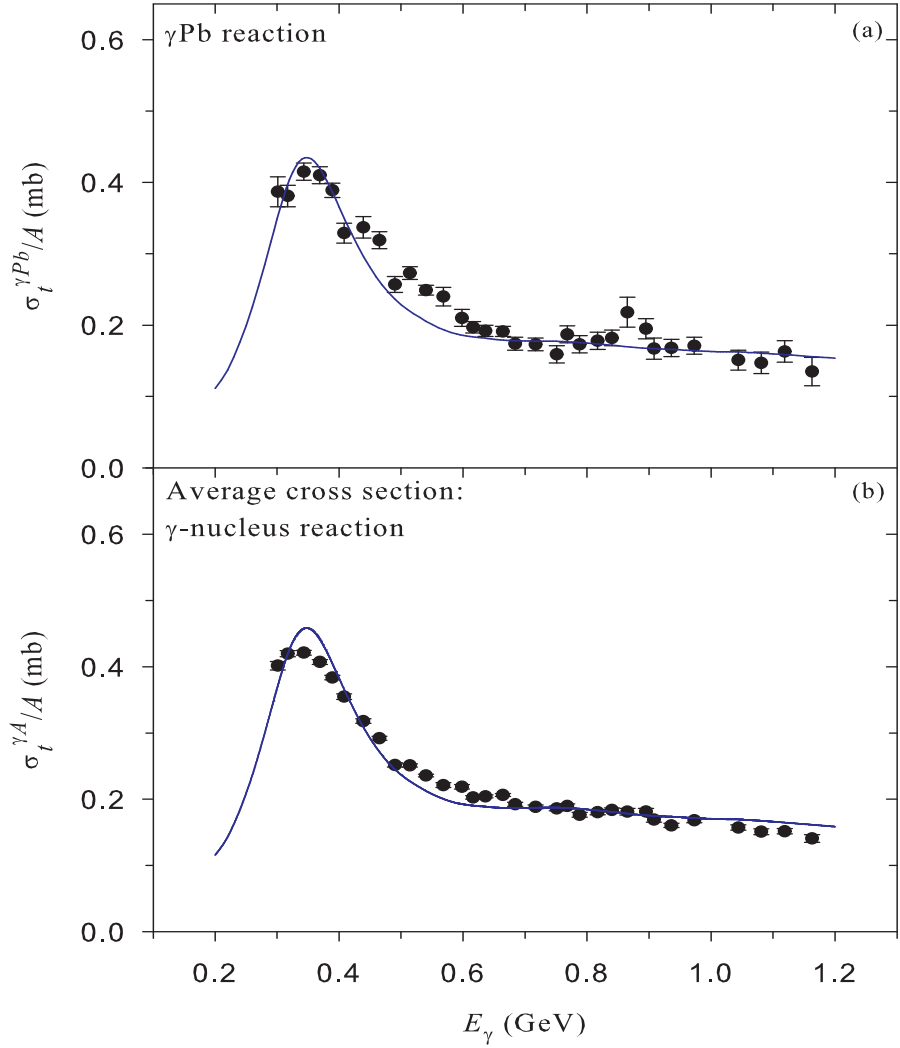,height=14.0 cm,width=12.0 cm}
}}
\caption{
(color online).
Upper part (a): 
The solid curve represent that in Fig.~\ref{XTgC} but for $^{208}$Pb nucleus. 
Lower part (b): 
The average of the cross sections calculated for $^{12}$C and $^{208}$Pb nuclei (denoted by the solid curve) is 
compared with the measured value. The data appearing in the figure are taken from Ref.~\cite {Bianchi}.
}
\label{XTgPbA}
\end{center}
\end{figure}

The cross section of the photoinduced reaction on the $^{208}$Pb nucleus has been evaluated using the 
Fermi-motion of the bound nucleon (as done for the $^{12}$C nucleus) and the resonance-nucleus interaction 
potential $V_{OR}$ in the calculation. The values of the in-medium scattering parameters (i.e., $\alpha_{RN}^*$ 
and $\sigma_t^{*RN}$) extracted from the photocarbon reaction, as discussed earlier, are used to determine  
$V_{OR}$. The calculated results (denoted by the solid curve) are compared with the data \cite{Bianchi} in 
Fig.~\ref{XTgPbA}(a). The average of the cross sections assessed for $^{12}$C and $^{208}$Pb nuclei, as shown 
by the solid curve in Fig.~\ref{XTgPbA}(b), agrees with the measured average cross section of the photonuclear 
reactions \cite{Bianchi}.

\section{Conclusions}

The cross section of the photonuclear reaction has been calculated to explore the scattering parameters of the 
resonances in the nucleus. The calculated results for the $^{12}$C nucleus are fitted with the data to extract 
the in-medium scattering parameters of the resonances. Using those parameters, the cross section of the 
photolead reaction is evaluated and compared that with the data. The average measured cross section of the 
photonuclear reactions is reproduced well by that of the calculated results.

\section{Acknowledgement}

The author thanks A. K. Gupta and S. M. Yusuf for their encouragement to work on the theoretical nuclear physics.







\newpage

\newpage

\newpage

\newpage

\newpage

\newpage

\end{document}